\documentclass[amsmath,superscriptaddress,amssymb,aps,twocolumn]{revtex4}
\usepackage{graphicx}
\usepackage{soul}
\usepackage[colorlinks=true,citecolor=blue,linkcolor=magenta]{hyperref}
\usepackage[usenames]{color}

\newcommand{\ket}[1]{\left| #1 \right>} 

\begin{document}

\title{
Controlling Correlated Tunneling and Superexchange Interactions \\
with AC-Driven Optical Lattices
}

\author{Yu-Ao Chen}
\thanks{These authors contributed equally to this work}
\author{Sylvain Nascimb\`ene}
\thanks{These authors contributed equally to this work}
\author{Monika Aidelsburger}
\author{Marcos Atala} 
\author{Stefan Trotzky}
\author{Immanuel Bloch}
\affiliation{Fakult\"at f\"ur Physik, Ludwig-Maximilians-Universit\"at, Schellingstrasse 4, 80799 M\"unchen, Germany}
\affiliation{Max-Planck-Institut f\"ur Quantenoptik, Hans-Kopfermann-Strasse 1, 85748 Garching, Germany}

\date{\today}

\begin{abstract} 
The dynamical control of tunneling processes of single particles plays a major role in science ranging from Shapiro steps in Josephson junctions to the control of chemical reactions via light in molecules. Here we show how such control can be extended to the regime of strongly interacting particles. Through a weak modulation of a biased tunnel contact, we have been able to coherently control single particle and correlated two-particle hopping processes. We have furthermore been able to extend this control to superexchange spin interactions in the presence of a magnetic-field gradient. We show how such photon assisted superexchange processes constitute a novel approach to realize arbitrary XXZ spin models in ultracold quantum gases, where transverse and Ising type spin couplings can be fully controlled in magnitude and sign.
\end{abstract}

\maketitle

The control of quantum tunneling of particles through a barrier using an oscillatory driving field lies at the heart of the interpretation of the so-called Shapiro steps observed in the $I$--$V$ characteristics of a biased Josephson junction under an applied radio-frequency field  \cite{shapiro1963josephson,barone1982physics}. Since then, other examples and applications of photon-assisted tunneling have emerged in several fields, such as the control of chemical reactions with coherent laser pulses  \cite{tannor1985control} or the observation of dynamic localization and absolute negative conductance in semiconductor superlattices  \cite{keay1995dynamic}. More recently the tunnel dynamics of single atoms in periodically-modulated optical lattices was investigated, showing in particular the possibility to revert the sign of the tunnel coupling for strong driving amplitudes  \cite{eckardt2005superfluid,lignier2007dynamical,kierig2008single,sias2008observation,eckardt2010frustrated,Struck11}. Periodically shaken optical lattices were also used as a spectroscopic tool for measuring the excitation spectrum of a superfluid Bose gas \cite{schori2004excitations} or extracting nearest-neighbor spin correlations in a fermionic Mott insulator \cite{greif2010probing}, as well as to study atomic transport in a quantum ratchet \cite{salger2009directed}.

In this article we investigate atom tunneling in an optical lattice of periodically modulated or `AC-driven' double-well potentials \cite{teichmann2009fractional,esmann2011fractional}. We first study the influence of atomic interactions on the dynamics of a single atom in the presence of the driving. This technique constitutes a precision spectroscopic tool for measuring interaction energies and inhomogeneities in optical lattices \cite{simon2011quantum}. We furthermore demonstrate the ability to control correlated tunneling processes, i.e. the co-tunneling of repulsively bound atom pairs \cite{fölling2007direct}. Moreover, we are able to drive superexchange interactions in the modulated double-well potentials, where the direct undriven exchange of spins is inhibited by an applied magnetic-field gradient. This result eventually leads to a novel proposal for implementing models of quantum magnetism with atomic gases in optical lattices \cite{duan2003controlling,garcía2003spin,kuklov2003counterflow,altman2003phase,lewenstein2007ultracold}. Indeed the generalization to a two-component atomic gas in a periodically-modulated  lattice realizes an XXZ spin model whose transverse and longitudinal couplings can be independently tuned by changing the strength of the driving or the amplitude of the magnetic-field gradient.

\begin{figure}[t!]
\includegraphics[width=\linewidth]{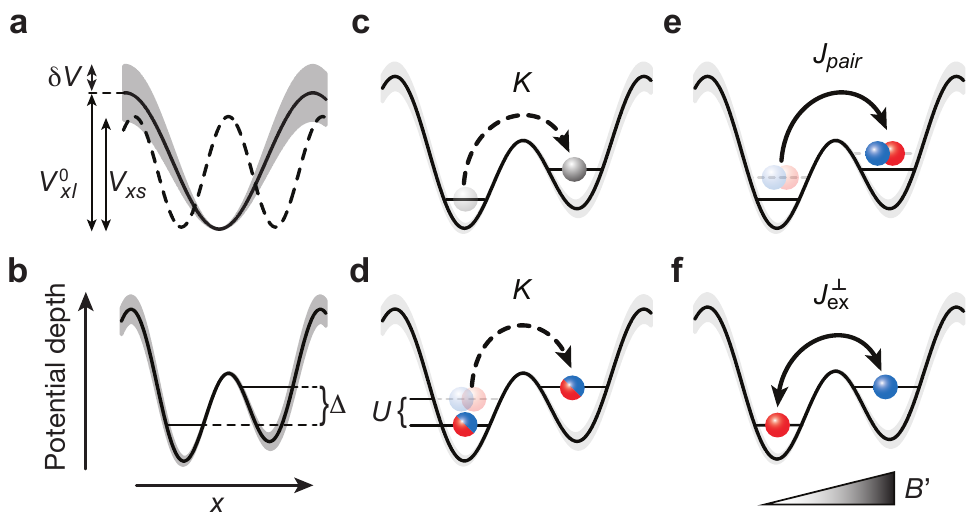}
\vspace{-0.5cm} \caption{AC-driven double-wells. \textbf{a,b}~Schematics of the superlattice potential used in our experiments. \textbf{c-f}~AC-driven tunneling processes studied in this work: \textbf{c} tunneling of a single atom, \textbf{d} tunneling of one atom out of an interacting atom pair, \textbf{e} correlated tunneling of an atom pair, and \textbf{f} spin exchange.\label{Fig_1}} 
\end{figure}


The physical system studied in our experiment consisted of an ensemble of $^{87}$Rb atoms held in a lattice of isolated double-wells \cite{fölling2007direct}. We first loaded a quasi-pure Bose-Einstein condensate of about $10^5$  atoms in the $\left|F=1,m_F=-1\right>$  Zeeman state into a 3D optical lattice formed by three retroreflected beams of laser light at the wavelengths $\lambda_{xl}=1534$~nm along $x$ direction (`long lattice'), $\lambda_y=844$~nm along $y$ and $\lambda_z=767$~nm along $z$. The final lattice depths were chosen so that the atomic sample was in the Mott insulating regime \cite{fisher1989boson,jaksch1998cold,greiner2002quantum} with a central core of two atoms per well and an outer shell of singly occupied sites. We then transferred all atoms to the $\left|F=1,m_F=0\right>$ state by a radio-frequency Landau-Zener adiabatic passage. Using microwave-dressed  spin-changing collisions \cite{widera2005coherent,gerbier2006resonant} we converted atom pairs in individual lattice wells into pairs with opposite magnetic moment, labeled as $\left|\uparrow\right> \equiv \left|F=1, m_F=-1\right>$ and $\left|\downarrow\right> \equiv \left|F=1, m_F=1\right>$.  In this process, single atoms remained in the $\left|F=1,m_F=0\right>$ Zeeman state.

\begin{figure*}[t!]
\includegraphics[width=\linewidth]{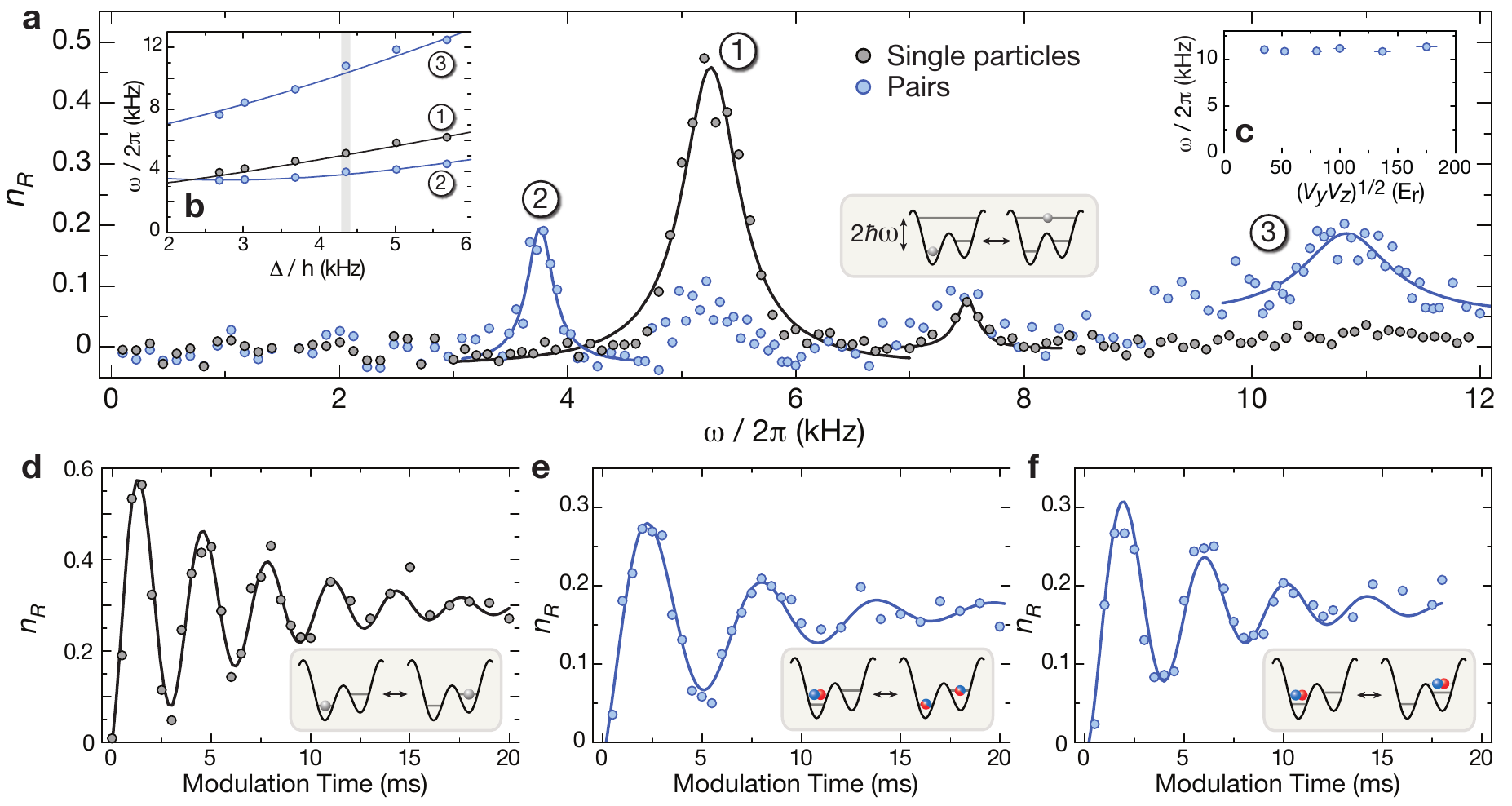}
\vspace{-0.5cm} \caption{Spectroscopic and coherent dynamical signals of AC-driven tunneling. \textbf{a} Fraction of atoms transferred to the right well $n_R$ as a function of the driving frequency $\omega/2\pi$ with a fixed modulation time $T=2.5$~ms (blue dots: atoms in $\left|\downarrow\right>$, black dots: atoms in $\left|F=1,m_F=0\right>$). The lattice parameters were $V_{xl}=35(1)~E_r^{xl}$, $V_{xs}=7.0(2)~E_r^{xs}$, $\phi=0.18(1)$~rad, $V_y=76(3)~E_r^y$, $V_z=77(3)~E_r^z$, and $\delta V=8.2(3)$~$E_r^{xl}$. 
Single atoms are resonantly transferred at $\omega/2\pi=5.2$~kHz, corresponding to $\hbar\omega=\sqrt{\Delta^2+4J^2}$. The dynamics of the atom pairs was monitored on the $\left|\downarrow\right>$ component: the resonant transfer at $\omega/2\pi=3.8$~kHz corresponds to the transfer of a single particle to the right well, while the one at $\omega/2\pi=10.8$~kHz corresponds to the driven co-tunneling of pairs. Atoms in $\left|\uparrow\right>$ component show the same behavior. The resonance at $\omega/2\pi=7.5$~kHz can be attributed to a two-photon transfer to the third Bloch band.  \textbf{b} Resonance frequencies $\omega/2\pi$ for the driven tunneling processes described above as a function of the tilt $\Delta$.
The solid lines corresponds to the prediction of the Hubbard Hamiltonian $(\ref{eq_Hubbard})$ with $J/h=1.3$~kHz and $U/h=2.9$~kHz.  The vertical line corresponds to the lattice configuration used in \textbf{a}. \textbf{c} Resonance frequencies $\omega/2\pi$ of the driven co-tunneling process as a function of the transverse lattice depth. The horizontal error bars represent a 3\% error in the lattice depth and the vertical error bars  represent the standard deviation of the fitted position of the co-tunneling resonance peak. \textbf{d-f} Time evolution of $n_R$ for the three resonances described above, together with fits using a damped sine wave. The measurements for \textbf{d} and \textbf{e} were performed with a larger modulation amplitude ($\delta V=16.4(5)$~$E_r^{xl}$ and $\delta V=10.2(3)$~$E_r^{xl}$, respectively).  \label{Fig_2}} 
\end{figure*}

An additional standing wave along the $x$ direction at $\lambda_{xs}\simeq\lambda_{xl}/2=767$~nm (`short lattice') was applied to create a periodic double-well potential $V(x)=V_{xl}\sin^2(k_lx)+V_{xs}\sin^2(2k_lx+\pi/2-\phi)$, where $k_l=2\pi/\lambda_{xl}$ (see Fig.~1a) and $\phi=0$ corresponds to a lattice of symmetric double wells. The relative phase $\phi$ and the lattice depths $V_{xl}$ and $V_{xs}$ could be independently controlled in real time by dynamically adjusting  the short-lattice wavelength $\lambda_{xs}$ and the laser intensities \cite{sebby2006lattice,fölling2007direct}.  
We express the depth of each lattice in units of its corresponding recoil energy $E_r^i=h^2/2m\lambda_i^2$, where $i=xs,xl,y,z$. Here, $m$ is the $^{87}$Rb atomic mass and $h=2\pi\hbar$ is Planck's constant.

The vibrational level splitting being much larger than the other relevant energy scales, the system can be described by a two-site Hubbard-like Hamiltonian,
%
%
\begin{eqnarray}
\hat{H}&=&-J\sum_{\sigma=\uparrow,\downarrow}\left(\hat{a}^\dagger_{L\sigma}\hat{a}_{R\sigma}^{\phantom{\dagger}}+\hat{a}^\dagger_{R\sigma}\hat{a}_{L\sigma}^{\phantom{\dagger}}\right)-\frac{\Delta}{2}\left(\hat{n}_{L}-\hat{n}_{R}\right)\nonumber\\
&+&\frac{U}{2}\left[\hat{n}_{L}(\hat{n}_{L}-1)+\hat{n}_{R}(\hat{n}_{R}-1)\right]\nonumber\\
&+&\frac{G}{2}(\hat{n}_{L\downarrow}-\hat{n}_{L\uparrow}-\hat{n}_{R\downarrow}+\hat{n}_{R\uparrow}),\label{eq_Hubbard}
\end{eqnarray}
where $J$ is the tunnel coupling, $U$ is the onsite interaction energy, and $\Delta$ is the potential tilt between the two sites. The last term represents a spin-dependent bias $G$ provided by an additional magnetic-field gradient along the $x$ direction. The operator $\hat{a}_{R(L)\sigma}$ annihilates a particle of spin $\sigma$ in the right (left) well, $\hat{n}_{R(L)\sigma}=\hat{a}_{R(L)\sigma}^\dagger\hat{a}_{R(L)\sigma}^{\phantom{\dagger}}$ is the corresponding number operator, and $\hat{n}_{R(L)}=\hat{n}_{R(L)\downarrow}+\hat{n}_{R(L)\uparrow}$ is the total atom number per well.

In addition to the static double-well potential, a time-periodic modulation was applied through a variation of the long-lattice depth $V_{xl}(t) = V_{xl}^0 + \delta V\cos(\omega t)$ (see Fig.~1a,b). The modulation introduces an additional term $\hat{K}\cos(\omega t)$ to the Hamiltonian $(\ref{eq_Hubbard})$ that couples left and right wells and can induce driven atom tunneling. The coupling operator $\hat{K}$ can be written as $\hat{K}=\sum_{\sigma}(-K\hat{a}^\dagger_{L\sigma}\hat{a}_{R\sigma}^{\phantom{\dagger}}-K^*\hat{a}^\dagger_{R\sigma}\hat{a}_{L\sigma}^{\phantom{\dagger}})$, where the matrix element $K=\delta V\int\mathrm{d}x\,w_L^*(x)\sin^2(k_{L}x)w_R(x)/2$ is calculated from the Wannier functions $w_L(x)$ and $w_R(x)$ in the left and right wells, respectively \cite{zhao1996rabi}. For simplicity we have omitted processes which do not directly induce atom tunneling,  such as terms proportional to $\hat{a}^\dagger_{L\sigma}\hat{a}_{L\sigma}^{\phantom{\dagger}}$.

\bigskip
\noindent\textbf{Transfer of a single atom and effect of interactions}\\
Let us first introduce the concepts of AC-driven tunneling for the simple case of single atoms (see Fig.~1c) \cite{kierig2008single}. Each of them was initially loaded in the left well of a tilted double-well potential ($\Delta=h\times4.3(2)$~kHz) by ramping up the short lattice with a phase of $\phi =0.18$~rad. We then applied a modulation with an amplitude $\delta V/V_{xl}=0.23$ during a time $T=2.5$~ms. At the end of the modulation, we measured the number of atoms in the right wells by transferring them to a higher Bloch band and subsequently performing a band-mapping sequence \cite{greiner2001exploring,sebby2006lattice,fölling2007direct}. As shown in the black dots of Fig.~2a, we observe a resonant transfer of atoms to the right wells when $\omega/2\pi=5.2(1)$~kHz.
This is in agreement with the difference $\Delta^\prime=\sqrt{\Delta^2+4J^2}=h\times5.0(2)$~kHz between the two lowest eigenenergies of the Hamiltonian $(\ref{eq_Hubbard})$, where the value $J = h \times 1.30(5)\,{\rm kHz}$ was obtained from an independent measurement of single-particle tunnel oscillations in a symmetric double-well configuration  ($\phi = 0$).

The resonance condition can be pictured using the Floquet formalism \cite{hänggi1998driven}. We denote the eigenstates of the Hamiltonian by $\left| i, j\right>$, where $i$ ($j$) indicates the occupation number in the first (second) eigenstate corresponding to the left (right) well, respectively. Then the states $\left|1,0\right>$ and $\left|0,1\right>$  are dressed by the modulation potential through the introduction of an effective photon number $n$ associated with a quantized energy $n\hbar\omega$. The resonance can thus be described as the level crossing between the states $\left|1,0;n\right>$ of energy $E_0=n\hbar\omega-\Delta^\prime/2$, and $\left|0,1;n-1\right>$ of energy $E_0+\Delta^\prime-\hbar\omega$. In this picture, the driven tunneling process is thus accompanied by the absorption of one photon.
As illustrated in Fig.~2d, varying the modulation time $T$ on resonance we observe Rabi oscillations between the states $\left|1,0\right>$ and $\left|0,1\right>$, showing that the AC-driven atom tunneling is a coherent process. As expected for a single-photon process, the Rabi frequency $\omega_R/2\pi$ is found to be proportional to the driving amplitude $\delta V$, with ${\omega_R}/({2\pi}\delta V)=24(1)$~Hz/$E_r^{xl}$, in agreement with the value $26(1)$~Hz/$E_r^{xl}$ obtained from a single-particle band structure calculation. The damping of the Rabi oscillations is well accounted for by the tilt inhomogeneities in our atomic sample that we measure through the width of the resonance (see  Appendix). These inhomogeneities are mainly caused by the external harmonic trapping potential superimposed to the lattice.

The scenario of single-particle tunneling becomes slightly modified if a second atom is present in the double-well (see Fig.~1d). For our preparation scheme, this situation was realized with the pair of atoms being in the spin states $\left|\uparrow\right>$ and $\left|\downarrow\right>$. At the beginning of the modulation both atoms were located on the left well, which we denote by $\left|\uparrow\downarrow,0;n\right>$. This experiment was performed in the absence of a magnetic-field gradient. Therefore the modulation symmetrically acted on both spin states and coupled the atom pair to the spin-symmetric triplet state $\ket{t_0;n-1}=(\left|\uparrow,\downarrow;n-1\right>+\left|\downarrow,\uparrow;n-1\right>)/\sqrt{2}$. As shown in Fig.~2a (blue dots), the resonance frequency $\omega/2\pi=3.8(1)$~kHz is shifted downwards with respect to the single-atom resonance. While in the initial state $\left|\uparrow\downarrow,0\right>$ the atoms are located on the same site and thus do maximally interact, they are essentially spatially separated in the final state, leading to an additional energy shift due to interactions. As shown in Fig.~2b, the measured resonance frequencies are well accounted for by the Hubbard Hamiltonian $(\ref{eq_Hubbard})$ in the whole parameter range $0<\Delta<5J$ ($J=h\times1.3$~kHz). Here, the on-site interaction $U=h\times2.9(1)$~kHz is obtained from a single-band calculation of the localized Wannier functions for $\Delta=0$ using a $s$-wave scattering length of $a_s=5.61$ nm. We observe that the linewidth of this resonance is smaller than the one of single atoms, which can be explained by smaller inhomogeneities in the tilt, since the spatial extent of the core of atom pairs is less wide than that of the outer shell of single atoms. In Fig.~2e, we show a measurement of the respective Rabi oscillations on resonance.

\bigskip
\noindent\textbf{Co-tunneling of an atom pair}\\
Interestingly, we also observe a resonant transfer of atoms to the right well for $\omega/2\pi=10.8(2)~\mathrm{kHz}$, which is about twice the resonance frequency for single atoms. 
We identify this resonance as a driven co-tunneling of  both atoms in a pair towards the state $\left|0,\uparrow\downarrow;n-1\right>$ \cite{teichmann2009fractional,esmann2011fractional} (see Fig.~1e). As expected for a transition between states with essentially the same interaction energy, we observe that the resonance frequency barely varies when the interaction energy $U$ is increased by increasing the transverse-lattice depths (see Fig.~2c). In addition, we directly probed the atom number distribution in the final state using spin-changing collisions after the modulation (see  Appendix). The obtained results show that the atoms indeed always tunnel to the right well as a repulsively bound atom pair \cite{winkler2006repulsively}, while for the resonance at $\omega/2\pi=3.8$~kHz only one atom is transferred. We observe that the linewidth of the co-tunneling resonance is larger than the one of the single-particle tunneling resonance. For the co-tunneling resonance condition $\hbar\omega\simeq2\Delta$ the effect of tilt inhomogeneities across the atomic sample is doubled with respect to the single-atom resonance.


\begin{figure}[t!]
\includegraphics[width=\linewidth]{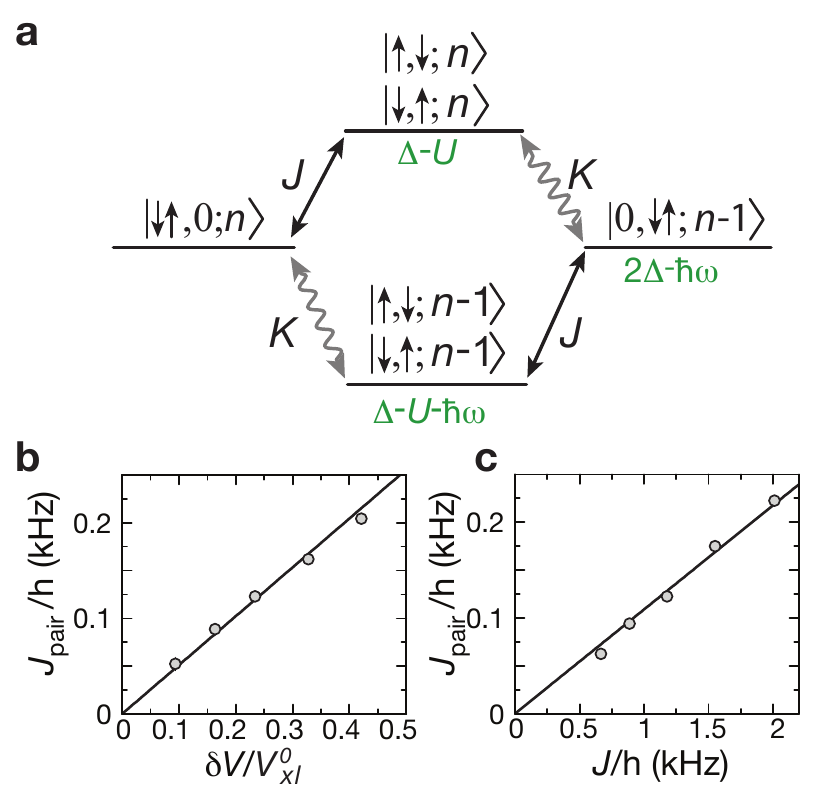}
\vspace{-0.5cm} \caption{Driven co-tunneling process. \textbf{a} Schematics of the levels involved in the co-tunneling of a repulsively bound pair. The states $\left|\uparrow\downarrow,0;n\right>$ and $\left|0,\uparrow\downarrow;n-1\right>$ are brought into resonance for $\hbar\omega=2\Delta$. The co-tunneling can then be understood as a second-order process mediated via virtual intermediate states, reached either by bare tunneling (coupling $J$) or AC-driven tunneling (coupling $K$) of single atoms. \textbf{b}  Co-tunnel coupling $J_{\mathrm{pair}}$ measured on resonance as a function of the modulation amplitude $\delta V$. The error bars representing the standard deviation of the Rabi oscillation frequency as obtained by a damped sine-wave fit are obscured by the data points. \textbf{c} Co-tunnel coupling $J_{\mathrm{pair}}$ measured as a function of the bare tunnel coupling $J$ which is adjusted by changing the short-lattice depth. The solid lines in \textbf{b} and \textbf{c} are linear fits to guide the eye. \label{Fig_3}} 
\end{figure}

The physical mechanism of this co-tunneling \cite{fölling2007direct} can be understood as a second-order process via off-resonant intermediate states virtually reached through single-atom tunneling processes -- either driven or undriven (see Fig.~3a). As an example, $\left|\uparrow\downarrow,0;n\right>$ is coupled to $\left|\uparrow,\downarrow;n\right>$ via the bare tunneling of the $\ket{\downarrow}$ particle (with a coupling $J$), then $\left|\uparrow,\downarrow;n\right>$ is coupled to $\left|0,\uparrow\downarrow;n-1\right>$ via AC-driven tunneling of the $\ket{\uparrow}$ particle (with a coupling $K$). Summing the contributions from all possible intermediate states, we obtain the pair tunneling coupling strength on resonance ($\hbar\omega=2\Delta$) within second-order perturbation theory  in the tunnel couplings,
\begin{equation}
J_{\mathrm{pair}}=2J K\left(\frac{1}{\Delta-U}+\frac{1}{-\Delta-U}\right).\label{Jpair}
\end{equation}
As shown in Fig.~3b, the resonant Rabi frequency of the co-tunneling process scales linearly with $K$ (proportional to the modulation amplitude $\delta V$). Varying the bare tunnel coupling $J$ by changing the short-lattice depth $V_{xs}$, we observe that $J_{\mathrm{pair}}$ also scales linearly with  $J$. A single-particle calculation predicts a variation of both $J$ and $K$ when varying $V_{xs}$, and thus in this approximation $J_{\mathrm{pair}}$ should not be proportional to $J$; however a numerical calculation of $J_{\mathrm{pair}}$ including five Wannier orbitals shows that couplings to higher bands due to atomic interactions lead to an effective linearization of the dependence of $J_{\mathrm{pair}}$ with $J$. Since these couplings tend to modify the AC-driven tunnel coupling $K$ with respect to the single-particle value, we directly measured the driven coupling strengths, for example by measuring the Rabi frequency of the $\left|\uparrow\downarrow,0;n+1\right>\rightarrow\left|t_0;n\right>$ process on resonance (see Fig.~2e). Equation (\ref{Jpair}) then predicts a Rabi frequency for the driven pair tunneling $2J_{\mathrm{pair}}/h=190(10)$~Hz, which is reasonably close to the measured value of $215(5)$~Hz (see Fig.~2f). 

\bigskip
\noindent\textbf{AC-driven superexchange interactions}\\
Having demonstrated the driven co-tunneling of an atom pair, we now apply the method of AC-driving to control the correlated tunneling process of  superexchange interactions. The superexchange of particles mediated via single particle to off-resonant intermediate states is the basic next-neighbor interaction mechanism in models of quantum magnetism arising in two-component Mott insulators \cite{auerbach1994interacting,duan2003controlling,garcía2003spin,kuklov2003counterflow,altman2003phase}.  Let us consider a one-dimensional Bose-Hubbard chain with two species of bosons. For large interactions $U\gg J$, the subspace of the Hilbert space with one atom per site is separated from the other states by the interaction energy $U$. However, neighboring spins can be exchanged through second-order tunneling processes \cite{duan2003controlling,garcía2003spin,trotzky2008time}, leading to an effective spin-spin interaction of the Heisenberg type $\hat{H}_{\mathrm{eff}}=-J_{\mathrm{ex}}\sum_{\left<i,j\right>}\mathbf{\hat{S}_i}\cdot\mathbf{\hat{S}_j}$. Here $J_{\mathrm{ex}}=4J^2/U$ is the superexchange coupling and the sum is made over pairs of neighboring sites. The spin operators are defined in Refs.  \cite{duan2003controlling,garcía2003spin,kuklov2003counterflow,altman2003phase}. In the presence of a magnetic-field gradient, the exchange of a pair of spins $\ket{\uparrow,\downarrow}\rightarrow\ket{\downarrow,\uparrow}$ is associated with an energy cost of $2G$. Therefore, superexchange processes are inhibited as soon as $G\gg J_{\mathrm{ex}}$ and the ground state corresponds to two spatially separated spin-polarized regions $\ket{\uparrow,\ldots,\uparrow,\downarrow,\ldots,\downarrow}$. By modulating the lattice potential at the resonance condition $\hbar\omega=2G$, however, it is possible to restore the resonant exchange of spins. In the dressed state picture, the system can then be mapped onto an ensemble of interacting spins, with the possibility to simulate an arbitrary XXZ model (see Methods)
\begin{equation}
\hat{H}_{\mathrm{eff}}=-\sum_{\left<i,j\right>}\left[J_\mathrm{ex}^\perp(\hat{S}_{i}^x\hat{S}_{j}^x+\hat{S}_{i}^y\hat{S}_{j}^y)+J_\mathrm{ex}^z\hat{S}_{i}^z\hat{S}_{j}^z\right]\label{XXZ}.
\end{equation} 
For example, in the case $J^2/U\ll G\ll U$ and for small driving amplitude, one obtains  $J_\mathrm{ex}^\perp\simeq8JK/U$ while the spin coupling in the $z$ direction is identical to the undriven case $J_\mathrm{ex}^z\simeq4J^2/U$. In particular it is possible to simulate with this system an Ising Hamiltonian for $K\ll J$ \cite{simon2011quantum}. As shown in Fig.~4, the couplings $J_\mathrm{ex}^\perp$ and $J_\mathrm{ex}^z$ and with them the anisotropy of the effective XXZ model can also be tuned by adjusting the values of $G$ and $U$, or by using a superlattice potential that lifts ever second site by an amount $\Delta$ in energy (see Supplementary Material). It then becomes possible to simulate pure Ising and XY models with ferromagnetic or anti-ferromagnetic interactions without the need to tune the spin-dependent on-site interaction energies.
%

\begin{figure}[t!]
\includegraphics[width=\linewidth]{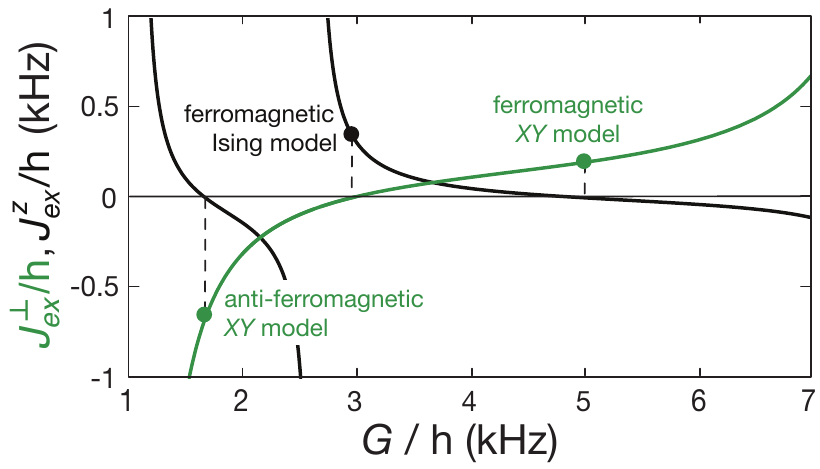}
\vspace{-0.5cm} \caption{Transverse and longitudinal superexchange couplings $J_\mathrm{ex}^\perp$, $J_\mathrm{ex}^z$ as a function of the magnetic-field gradient $G$. For this plot the single-atom tunnel couplings are taken to be $J=h\times0.5$~kHz and $K=h\times0.6$~kHz, the tilt induced by a superlattice potential is $\Delta=h\times3.4$~kHz and the onsite interaction energy is $U=h\times4.5$~kHz. The calculations are performed using perturbation theory up to quadratic order in $J$ and $K$ (see Methods).  \label{Fig_3A}} 
\end{figure}
%

In order to probe AC-driven superexchange interactions in our system of double-wells (see Fig.~1f), we loaded the atomic spin pairs in symmetric double-wells in the presence of a magnetic-field gradient along the $x$ direction. Its strength $G/h=1.2(1)$~kHz was measured from the shift of the single-particle resonance of $\left|\downarrow\right>$ atoms. The degeneracy between $\left|\uparrow,\downarrow\right>$ and $\left|\downarrow,\uparrow\right>$ was hence lifted and the atoms were occupying the ground state $\left|\uparrow,\downarrow\right>$. We carried out the modulation spectroscopy as in the previous cases and a typical spectrum is displayed in Fig.~5 for a tilt of $\Delta=h\times8.4$~kHz. We observe two kinds of resonances in this spectrum. 
First for $\omega/2\pi=4.5(2)$~kHz and $\omega/2\pi=13.0(2)$~kHz only the atoms in one of the spin states are transferred: the transfer of $\left|\downarrow\right>$ atoms to the left site occurs at $\hbar\omega=\left|-\Delta+U+G\right|$ and the transfer of $\left|\uparrow\right>$ particles to the right site at $\hbar\omega=\left|\Delta+U+G\right|$. For these measurements, the on-site interaction energy was $U\simeq h\times3.4(3)$~kHz. For the second type of resonances both spin states are transferred simultaneously in an AC-driven superexchange process.

%
\begin{figure}[t!]
\includegraphics[width=\linewidth]{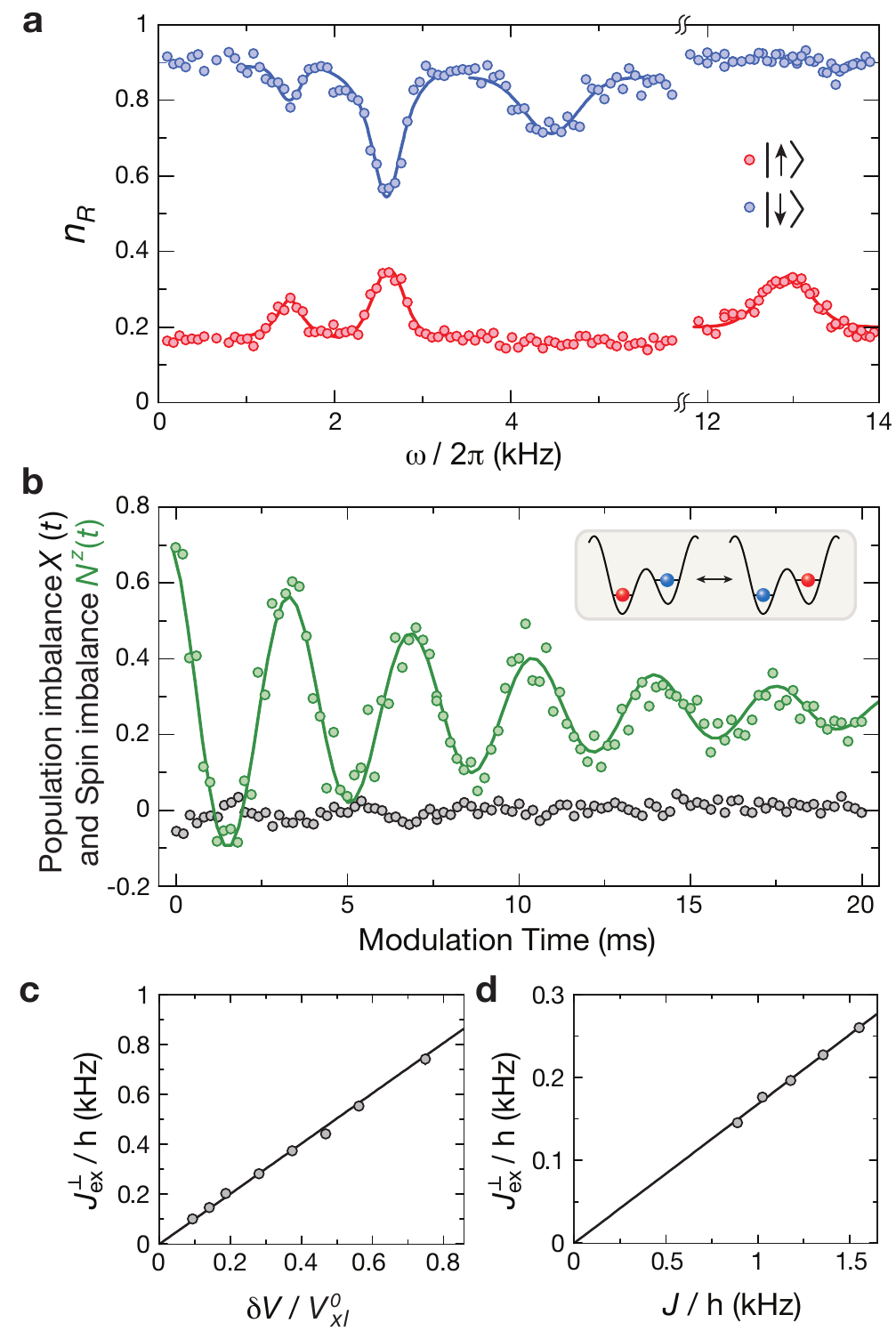}
\vspace{-0.7cm} \caption{Driven superexchange process. \textbf{a} Fraction of atoms $n_R$ in the right well as a function of the modulation frequency $\omega/2\pi$ (blue dots: spin $\ket{\uparrow}$, red dots: spin $\ket{\downarrow}$). We observe a resonant transfer of both spin states for $\omega/2\pi=2.6$~kHz corresponding to a single-photon-driven superexchange process. The resonance at $\omega/2\pi=1.5$~kHz corresponds to a spin exchange driven by the absorption of two photons in total. In addition, we observe the single-particle tunneling resonances at $\omega/2\pi=4.5$~kHz for the $\ket{\downarrow}$ atoms and $\omega/2\pi=13$~kHz for the $\ket{\uparrow}$ atoms. For these data the lattice depths were $V_z=192(6)$~$E_r^z$, $V_y=142(5)$~$E_r^y$ for the transverse lattices and $V_{xl}=35(1)$~$E_r^{xl}$, $V_{xs}=7.0(2)$~$E_r^{xs}$ for the superlattice, and the superlattice phase was $\phi=0.35(1)$~rad. \textbf{b} Time evolution of the mean population imbalance $X(t)$ and spin imbalance $N^z(t)$ at the superexchange resonance $\omega/2\pi=2.6$~kHz, fitted with a damped sine wave. \textbf{c,d} Driven superexchange coupling $J_\mathrm{ex}^\perp$ measured as a function of (\textbf{c}) the modulation amplitude $\delta V$  and (\textbf{d}) the tunnel coupling $J$, together with linear fits to the data as a guide to the eye. The error bars representing the standard deviation of the Rabi oscillation frequency as obtained by a damped sine-wave fit are obscured by the data points.  \label{Fig_4}} 
\end{figure}

%
Let us focus on the resonance occurring at $\omega/2\pi=2.6(1)$~kHz, which corresponds to the resonance condition $\hbar\omega\simeq2G$ for the driven superexchange. As shown by the Rabi oscillation in Fig.~5b, on resonance both spin states coherently tunnel between left and right wells in a correlated manner. At any time the population imbalance between both wells $X(t)=(n_{L}-n_{R})/2$ remains equal to 0, as expected for a spin-exchange process. Our measurement scheme corresponds to a projection of the quantum state on left and right wells; due to the finite tunnel coupling $J$ the actual eigenstates involved in the superexchange process are not fully localized. Taking also into account the coupling to higher bands, we calculate that for our trap parameters the maximum value of the mean spin imbalance or N\'eel order parameter $N^z=(n_{\uparrow L}-n_{\uparrow R}-n_{\downarrow L}+n_{\downarrow R})/2$ amounts to 0.8, close to the measured value $N^z(t=0)=0.7$. The superexchange oscillation occurs at a rate $J_\mathrm{ex}^\perp/h=560(20)$~Hz and is damped with a $1/e$ time $\tau=9(1)$~ms, most likely due to inhomogeneities in the magnetic-field gradient. 
Due to the latter, the Rabi oscillation is detuned for part of the atomic sample, leading to a non-zero asymptotic value of $N^z(t\gg\tau)=0.26(2)$. Finally, we show in Fig.~5c that the driven superexchange coupling $J_\mathrm{ex}^\perp$ scales linearly with the driving amplitude $\delta V$, as expected for a single-photon assisted correlated tunneling process.

The driven superexchange coupling can be viewed as a second-order process in which one particle virtually tunnels onto the second particle before the latter tunnels in the other direction (see Fig.~6a). Similarly to the pair tunneling process, one virtual tunneling spontaneously occurs with the coupling $J$ while the other one is driven by the absorption of one photon of frequency $\hbar\omega=2G$. Taking into account the four possible intermediate states, we obtain the superexchange coupling with second-order perturbation theory: 
\begin{eqnarray}
 J_\mathrm{ex}^\perp&=&2JK\left(\frac{1}{\Delta+U+G}+\frac{1}{\Delta+U-G}\right.\nonumber\\
             &+&\left.\frac{1}{-\Delta+U+G}+\frac{1}{-\Delta+U-G}\right).
\label{Jex}
\end{eqnarray} 
%

\begin{figure}[t!]
\includegraphics[width=\linewidth]{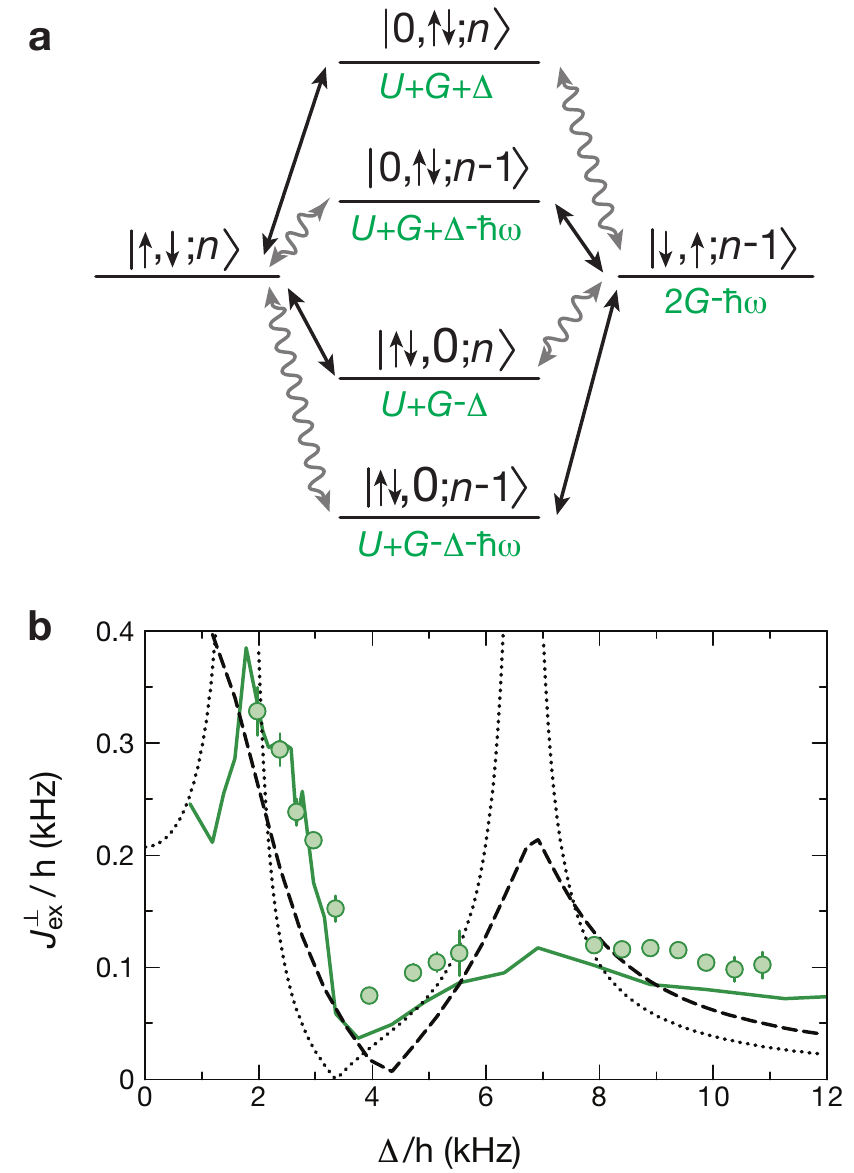}
\vspace{-0.5cm} \caption{Control of superexchange interactions with the potential tilt $\Delta$. \textbf{a} Schematics of the driven superexchange process. The superexchange interaction between the states $\left|\uparrow,\downarrow;n\right>$ and $\left|\downarrow,\uparrow;n-1\right>$ is mediated by bare tunneling and photon-assisted couplings towards four virtual intermediate states and is resonant for $\hbar\omega=2G$. \textbf{b} Strength of the driven superexchange interaction $J_\mathrm{ex}^\perp$ as a function of the tilt $\Delta$. We observe a resonant enhancement of the coupling when $\Delta$ approaches $U-G\simeq h\times1.7$~kHz, the virtual state  $\left|\uparrow\downarrow,0;n\right>$ becoming closer to resonance. The experimental data (open circles) is compared with three different predictions: second-order perturbation theory given by equation (\ref{Jex}) (dotted line), the non-perturbative solution of the Hamiltonian restricted to the six-level structure pictured in \textbf{a} (dashed line), and a numerical calculation including the effect of the coupling to higher bands (solid line). The error bars represent the standard deviation of the Rabi oscillation frequency as obtained by a damped sine-wave fit.  \label{Fig_5}} 
\end{figure}

%
Equation (\ref{Jex}) shows that by tuning the tilt $\Delta$ it is possible to resonantly enhance the superexchange coupling strength around the values $\Delta=U\pm G$, as well as to suppress  it for $\Delta=\sqrt{\left|U^2-G^2\right|}$. The measured variation of  $J_\mathrm{ex}^\perp$ with $\Delta$ is shown as open circles in Fig.~6b. In order to widen the validity range of second-order perturbation theory (set by the conditions $J,K\ll|\Delta\pm U\pm G|$), we increased the depth of the short lattice to 10~$E_r^{xs}$ so that the tunnel coupling $J$ was reduced to 0.5~kHz. For this lattice configuration, the resonances correspond to $\Delta=U+G=h\times6.7(2)$~kHz and $\Delta=U-G=h\times1.7(2)$~kHz, and $ J_\mathrm{ex}^\perp$ is predicted to cancel for $\Delta=h\times3.3(2)$~kHz. By decreasing the tilt from 3.9~kHz to 2~kHz we observe a resonant enhancement of the driven superexchange coupling over a factor of 5 (see Fig.~6b). While equation (\ref{Jex}) gives a qualitative description of our measurements (dotted line), we obtain a better quantitative agreement by including virtual transitions to higher bands (solid line, see  Appendix). 
%
%

Finally, we note that the resonance observed for $\omega/2\pi=1.50(5)~\mathrm{kHz}\simeq G/h$ also corresponds to an AC-driven superexchange process. Contrary to the previous case, we observe that the coupling  $J_\mathrm{ex}^\perp$ varies quadratically with the modulation amplitude $\delta V$, showing that the spin exchange is driven in that case by the absorption of two photons (see  Appendix). 

To conclude, we have shown that periodically-modulated optical superlattices provide a new tool for controlling (correlated) atom tunneling in an optical lattice. It can be used as a spectroscopic tool to measure the parameters of the underlying Hubbard model, and its generalization e.g. to the case of a pair of coupled many-body systems such as 1D gases could give a direct access to their spectral function. Moreover, extending the driving to a full lattice, the method provides a new degree of control of effective spin-spin interactions,  allowing one to access XXZ models with arbitrary anisotropy by merely tuning the magnetic-field gradient and the driving amplitude. To adjust $J_{ex}^\perp$ and $J_{ex}^z$ independently in the undriven case one would need to tune the spin-dependent on-site interaction energy \cite{duan2003controlling,garcía2003spin,kuklov2003counterflow,altman2003phase}. This can only be achieved by either using near-resonant spin-dependent lattices which will cause strong heating due to scattering of lattice photons, or by the use of Feshbach resonances which might be practically unavailable for many atomic species. The approach presented here does not suffer from these problems and can thus be conveniently applied to e.g. study the phase diagram of the XXZ model \cite{duan2003controlling} or to simulate the dynamics of XXZ spin chains \cite{Barmettler09,Barthel09} with ultracold atoms over a wide range of parameters.

\bigskip
\noindent\textbf{Methods}\\
\noindent\textbf{Mapping to an effective XXZ spin model.}
Let us consider a simple 1D chain with one atom per site. A magnetic-field gradient is applied along the longitudinal direction, leading to a spin-dependent tilt $G$. The exchange of neighboring atoms with opposite spins costs an energy $2G$. By modulating the lattice potential at the frequency $\omega/2\pi=2G/h$, this process is made resonant  with an effective superexchange coupling 
\begin{equation}
J_\mathrm{ex}^\perp=8JK\frac{U}{U^2-G^2},\label{eqJex}
\end{equation}
given by equation (\ref{Jex}) with $\Delta=0$. In the dressed state picture, all states with a given total atom number per spin $N_\sigma$ ($\sigma=\uparrow,\downarrow$) are degenerate when the photon number compensates the magnetic-field gradient:
\begin{equation}
n\hbar\omega=\sum_{i}[n_{i\downarrow}-n_{i\uparrow}]G.\label{nphotons}
\end{equation}
In the following we drop the photon number, assuming that its value is given by equation (\ref{nphotons}).
Virtual processes such as $\ket{\ldots,\uparrow,\downarrow,\ldots}\leftrightarrow\ket{\ldots,0,\uparrow\downarrow,\ldots}\leftrightarrow\ket{\ldots,\uparrow,\downarrow,\ldots}$ lead to a lift of the degeneracy between states with a given spin polarization. These energy shifts can be calculated using second-order perturbation theory and can be recast as an Ising interaction between neighboring sites
$\Delta E=-\sum_{\left<i,j\right>}J_\mathrm{ex}^zS_{i}^zS_{j}^z$, where $S_{i}^z=(n_{i\uparrow}-n_{i\downarrow})/2$ is the $z$-component of an effective spin $1/2$, and
\begin{equation}
J_\mathrm{ex}^z=\frac{4J^2}{U}\frac{U^2}{U^2-G^2}\label{eqJz}
\end{equation}
is the effective Ising coupling of this spin Hamiltonian (see  Appendix). Here we have assumed that the scattering length describing the collisions of a pair of atoms does not depend on their internal states, which is a good approximation for the case of $^{87}$Rb atoms \cite{van2002interisotope}.

In the low-gradient limit $G\ll U$, the Ising coupling is given by $J_\mathrm{ex}^z=4J^2/U$, while for $G\gg U$ one obtains $J_\mathrm{ex}^z=-4J^2U/G^2$. The sign of $J_\mathrm{ex}^z$ can be tuned negative by choosing $G>U$, allowing to simulate both ferromagnetic and antiferromagnetic cases. Combining equations (\ref{eqJex}) and (\ref{eqJz}), we obtain the effective XXZ model
\[
\hat{H}_{\mathrm{eff}}=-\sum_{\left<i,j\right>}\left[J_\mathrm{ex}^\perp(\hat{S}_{i}^x\hat{S}_{j}^x+\hat{S}_{i}^y\hat{S}_{j}^y)+J_\mathrm{ex}^z\hat{S}_{i}^z\hat{S}_{j}^z\right],
\]
where  $\hat{S}_{i}^x=(\hat{a}_{i\uparrow}^\dagger\hat{a}_{i\downarrow}^{\phantom{\dagger}}+\hat{a}_{i\downarrow}^\dagger\hat{a}_{i\uparrow}^{\phantom{\dagger}})/2$, $\hat{S}_{i}^y=(\hat{a}_{i\uparrow}^\dagger\hat{a}_{i\downarrow}^{\phantom{\dagger}}-\hat{a}_{i\downarrow}^\dagger\hat{a}_{i\uparrow}^{\phantom{\dagger}})/2i$ are the transverse effective spin operators. 

Note that this scenario can directly be extended to two- and three-dimensional lattices, where the gradient field is applied along the respective diagonal direction.

We would like to thank Simon F\"olling and Christoph Gohle for their help in building up the experiment, and Christoph Weiss for stimulating discussions. This work was supported by the DFG (FOR635, FOR801), the EU (STREP, NAMEQUAM, Marie Curie Fellowship to S.N.), and DARPA (OLE program).


\section*{Appendix}

\renewcommand{\thefigure}{A\arabic{figure}}
 \setcounter{figure}{0}
\renewcommand{\theequation}{A.\arabic{equation}}
 \setcounter{equation}{0}
 \renewcommand{\thesection}{A.\Roman{section}}
\setcounter{section}{0}

\section{Probing the atom number distribution using spin-changing collisions}
\begin{figure}[t!]
\includegraphics[width=\linewidth]{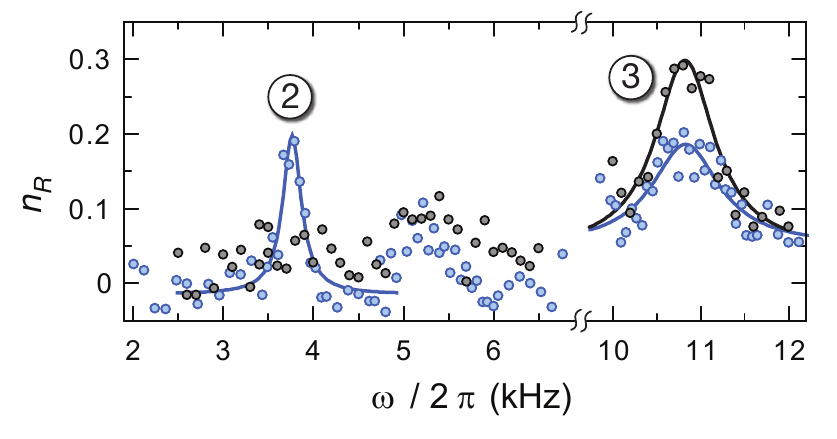}
\vspace{-0.5cm} \caption{Fraction of atoms $n_R$ transferred to the right well for $\left|\downarrow\right>$ atoms as a function of the modulation frequency $\omega/2\pi$. The lattice parameters are identical to the ones indicated in the legend of Fig.~2 in the main article. For the data in blue spin-changing collisions are performed before the modulation, and we observe two resonant transfers of atoms at $\omega/2\pi=3.8$~kHz and $\omega/2\pi=10.8$~kHz. For the data in black, the spin-changing collisions are performed after the modulation in order to test whether the final state consists of one atom per well or whether both atoms lie on the same well.  The absence of a resonance peak at $\omega/2\pi=3.8$~kHz shows that for this tunneling process only one atom is transfered to the right side.  \label{Fig_S1}} 
\end{figure}

As shown in Ref.  \cite{gerbier2006probing}, spin-changing collisions (SCC) can be used to measure the fraction of atom pairs in an optical lattice. Here, we use this technique to identify the final states of the AC-driven tunneling processes shown in Fig.~2a of the main article. We recall that SCC were used before the modulation to transfer atom pairs into the Zeeman states $\left|\uparrow\right>$ and $\left|\downarrow\right>$. In this spectrum, the atoms in $\left|\downarrow\right>$ are found to be resonantly transferred for $\omega/2\pi=3.8$~kHz and $\omega/2\pi=10.8$~kHz, and we want to show that the first resonance corresponds to the transfer of a single atom while for the second one both atoms tunnel to the right well at once. For this purpose we did not perform the SCC before the lattice modulation, but froze out all tunneling at the end of the driving and only then used the SCC to convert atom pairs into pairs of $\left|\uparrow\right>$ and $\left|\downarrow\right>$. As shown in Fig.~\ref{Fig_S1}, the resonance at $\omega/2\pi=10.8$~kHz is essentially unchanged, showing that the final state for this process consists of two particles in the right well, as expected for the co-tunneling process. On the contrary, the resonance at $\omega/2\pi=3.8$~kHz does not appear in the spectrum for atoms in state $\left|\downarrow\right>$, indicating that the final state of this process cannot undergo SCC, i.e. it is made of one particle on each site of the double-well potential.

\bigskip
\section{Effect of the coupling to higher bands}
The quantitative understanding of the amplitude of superexchange interactions requires to take the coupling to higher bands into account. We solve the two-body Hamiltonian numerically within the Hilbert space restricted to the lowest five Bloch bands of the optical superlattice. The eigenstates are expanded within the basis of the respective Wannier functions which are approximated by Bloch functions truncated to one lattice period and renormalized. The validity of this approximation relies on the large depth of the long lattice which suppresses tunneling between double wells. We include the effect of a magnetic-field gradient by adding to the lattice potential a periodic sawtooth potential that coincides with the linear $B$-field potential inside one double-well. We thus obtain a set of Wannier functions for each spin component. The effect of interactions is included by calculating all matrix elements of the contact interaction Hamiltonian  in the basis of the Wannier functions. After diagonalizing the whole Hamiltonian, we calculate the superexchange coupling as the matrix element of $\delta V\sin^2(k_{L}x)$ between the relevant eigenstates and obtain the solid curve plotted in Fig.~6 in the main article. 

\bigskip
\section{Width of the single-particle tunneling resonance}
While for large modulation amplitudes $\delta V$ the width of the single-particle tunneling resonance is essentially equal to the resonant Rabi frequency, it is limited for small modulation amplitudes by the inhomogeneities in the tilt $\Delta$ across the extent of the atomic sample. For trap parameters corresponding to Fig.~2 in the main article and for an atom number $N\simeq2\times10^4$, we measure a minimal width (FWHM) of the resonance peak of 0.35(10)~kHz. Here, the modulation time was $T=3.5$~ms, corresponding to a Fourier-limited width of 0.2~kHz. The most likely source of inhomogeneities stems from the Gaussian intensity profiles of the lattice beams (waist $\simeq125~\mu$m). First, the intensity variation  of the short and long lattices along the $x$ direction leads to an inhomogeneity of the depths $V_{xs}$ and $V_{xl}$ and thus of the superlattice potential. Second, the lattices along the transverse directions provide an additional overall harmonic confinement $\frac{1}{2}m\omega_x^2x^2$ along the $x$ direction that is superimposed to the superlattice and locally deforms the double-well potentials. For the parameters of Fig.~2 in the main article, we estimate a harmonic confinement  along the $x$ direction of $\omega_x/2\pi\simeq80$~Hz. Assuming that the atom number distribution is the one of a $T=0,\,J=0$  Mott insulator with one atom per double-well, we calculate the inhomogeneity in $\Delta$ due to both effects, and obtain a resonance width of 235~Hz, reasonably close to the measured value. In addition to the external confinement, the inhomogeneities in transverse directions due to any misalignment of the superlatice beams would further increase the width of the respective peak.


\begin{figure}[t!]
\includegraphics{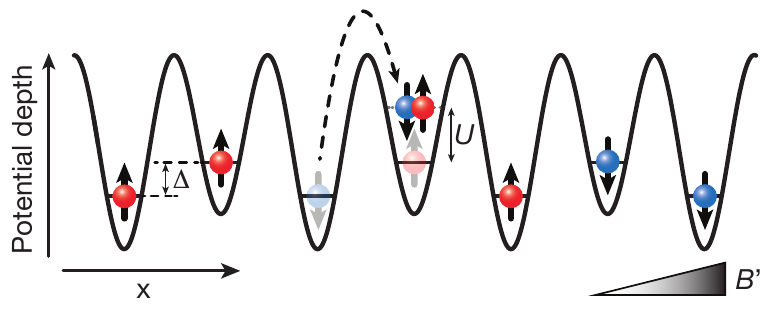}
\vspace{0.cm} \caption{Schematic representation of a spin chain in the 1D superlattice. The dashed arrow denotes the virtual hopping of a spin $\ket{\downarrow}$ atom on a neighboring site initially occupied by a spin $\ket{\uparrow}$ atom. For this particular process the energy difference between the final and initial states is $\Delta+U-G$.  \label{Fig_S2}} 
\end{figure}

\bigskip
\section{Mapping to an effective XXZ spin model}
In this section we describe how to calculate the lift of degeneracy of the single occupancy subspace due to single-particle virtual hopping, and map it onto an effective spin model. For simplicity we present the case of a 1D chain of atoms, held in an optical superlattice  with every second site raised in energy by $\Delta$ and in the presence of a gradient $G$ (see Fig.~\ref{Fig_S2}). The extension to two- and three-dimensional systems is straightforward. 

Let us consider the effect of the single-particle hopping  process of $\ket{\ldots,\downarrow,\uparrow,\ldots}\leftrightarrow\ket{\ldots,0,\downarrow\uparrow,\ldots}$ pictured in Fig.~\ref{Fig_S2}, that occurs with a coupling $J$ and is detuned by an energy $\Delta+U-G$. Within second-order perturbation theory, this virtual process
leads to an energy shift $-J^2/(\Delta+U-G)$ of the initial state.  Together with the virtual process in which the right particle hops on the left particle, one obtains the energy shift associated with  this atom pair: $-J^2[1/(\Delta+U-G)+1/(-\Delta+U-G)].$ Virtual hopping can also be driven by the absorption or emission of one photon, leading to an additional energy shift
\begin{eqnarray}
&&-K^2[1/(\Delta+U-G+\hbar\omega)+1/(-\Delta+U-G+\hbar\omega)\nonumber\\
&&+1/(\Delta+U-G-\hbar\omega)+1/(-\Delta+U-G-\hbar\omega)].\nonumber
\end{eqnarray} 

Similarly, each pair of neighboring atoms contributes to the energy shift with a total value $\delta E_{\sigma\sigma'}^\alpha$ that depends on the spin configuration ($\sigma$ for the left spin, $\sigma'$ for the right spin) and on whether the left particle is on a high well ($\alpha=+$) or not ($\alpha=-$). The total energy shift can then be written as
\begin{equation}
\delta E=\sum_{\sigma,\sigma',\alpha}N_{\sigma\sigma'}^\alpha\delta E_{\sigma\sigma'}^\alpha,\label{deltaE1}
\end{equation}
where $N_{\sigma\sigma'}^\alpha$ is the number of pairs of neighbor atoms in the configuration $\sigma,\sigma',\alpha$. The numbers of domain walls $\uparrow\downarrow$ and  $\downarrow\uparrow$ cannot differ by more than one, hence in the thermodynamic limit $N_{\uparrow\downarrow}=N_{\downarrow\uparrow}$, where $N_{\sigma\sigma'}=\sum_{\alpha}N_{\sigma\sigma'}^\alpha$. Moreover, a direct calculation shows that the energy shift does not depend on $\alpha$ for a pair of opposite spins. This allows us to replace in (\ref{deltaE1}) the terms with $\sigma\neq\sigma'$ by $(N_{\uparrow\downarrow}+N_{\downarrow\uparrow})(\delta E_{\uparrow\downarrow}^++\delta E_{\downarrow\uparrow}^+)/2$. Similar relations when considering the pairs of identical spins finally lead to (up to a constant)
\begin{equation}
\delta E=(N_{\uparrow\uparrow}+N_{\downarrow\downarrow})\frac{\delta E_{\uparrow\uparrow}^++\delta E_{\downarrow\downarrow}^+}{2}+(N_{\uparrow\downarrow}+N_{\downarrow\uparrow})\frac{\delta E_{\uparrow\downarrow}^++\delta E_{\downarrow\uparrow}^+}{2}.\label{deltaE2}
\end{equation}
Equation (\ref{deltaE2}) is nothing but the energy of a spin-$1/2$ chain with an Ising coupling $J_\mathrm{ex}^z=\delta E_{\uparrow\downarrow}^++\delta E_{\downarrow\uparrow}^+-\delta E_{\uparrow\uparrow}^+-\delta E_{\downarrow\downarrow}^+$. The transverse component $J_\mathrm{ex}^\perp$ is provided by the photon-assisted hopping and is given by equation (4) in the main article.

\begin{figure}[t!]
\includegraphics[width=\linewidth]{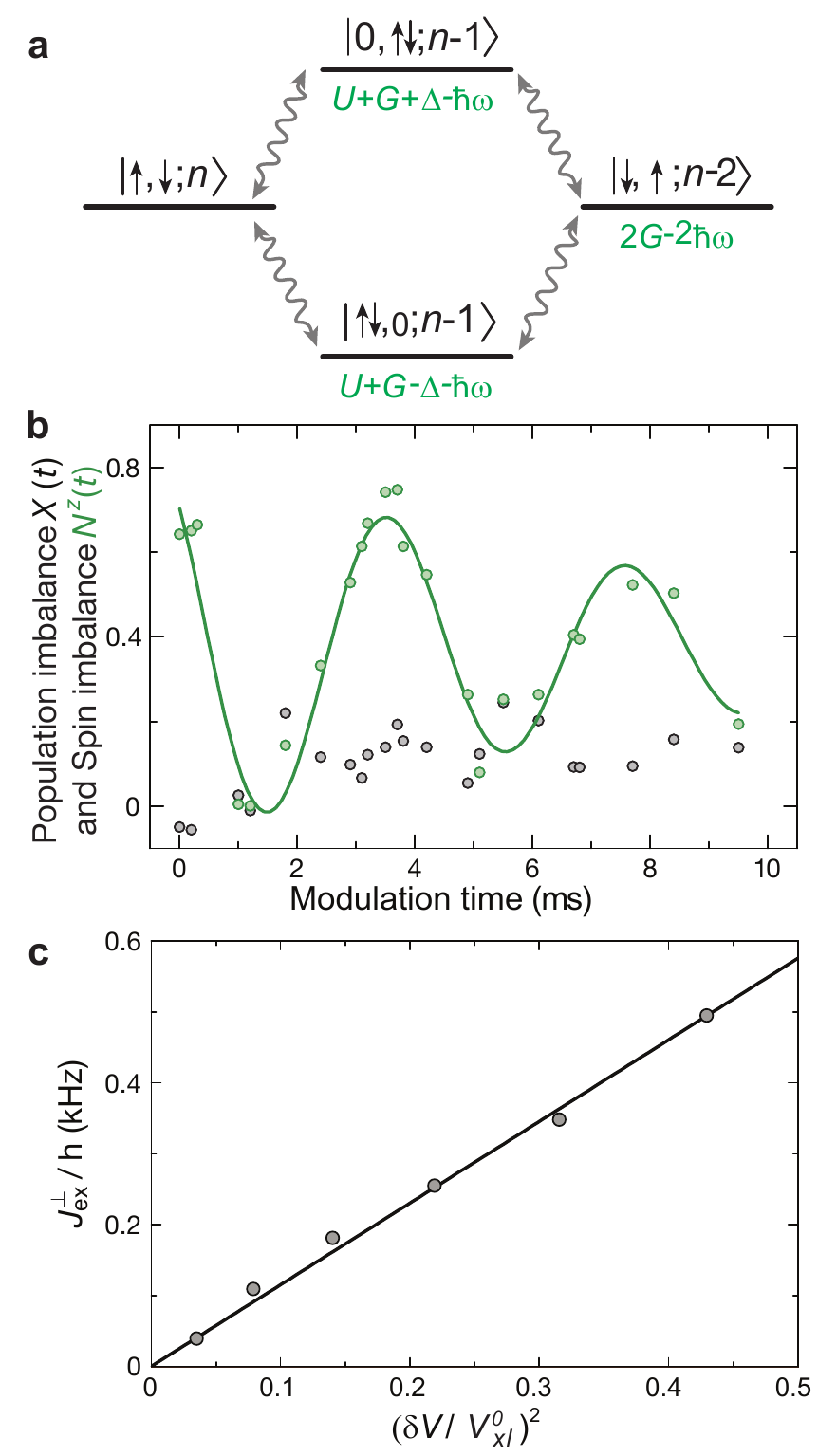}
\vspace{-0.5cm} \caption{Superexchange process for $\omega/2\pi=G/h$. \textbf{a} The superexchange interaction between the states $\left|\uparrow,\downarrow;n\right>$ and $\left|\downarrow,\uparrow;n-2\right>$ is mediated by photon-assisted couplings to virtual intermediate states and is resonant for $\hbar\omega=G$.  \textbf{b} Time evolution of the mean population imbalance $X$ and spin imbalance $N^z$ at the superexchange resonance $\omega/2\pi=1.5$~kHz, fitted with a damped sine wave. \textbf{c} Driven superexchange coupling $J_\mathrm{ex}^\perp$ measured as a function of the square of the modulation amplitude $\delta V$ together with a linear fit as a guide to the eye. \label{Fig_S3}} 
\end{figure}

\bigskip
\section{Two-photon-driven superexchange process}

While the energy cost of a spin exchange in the presence of a magnetic-field gradient is $2G$, we observe a resonant transfer of both spin states for $\omega/2\pi=G/h$ (see Fig.~5a in the main article). In addition, as shown in Fig.~\ref{Fig_S3}, on resonance the tunnel dynamics for both spin components is perfectly correlated, indicating that this process corresponds to a spin exchange. Its coupling strength is found to vary quadratically with the modulation amplitude (see Fig.~\ref{Fig_S3}), which shows that this superexchange process is driven by the absorption of two photons. The condition for a resonant spin exchange thus reads $2\hbar\omega=2G$, which explains the position of the resonance.


\begin{thebibliography}{10}
\providecommand{\url}[1]{\texttt{#1}}
\providecommand{\urlprefix}{URL }
\providecommand{\eprint}[2][]{\url{#2}}

\bibitem{shapiro1963josephson}
Shapiro and S., Phys. Rev. Lett. \textbf{11}, 80 (1963).


\bibitem{barone1982physics}
A.~Barone and G.~Paterno, \emph{{Physics and applications of the Josephson
  effect}}.
\newblock Wiley (1982).

\bibitem{tannor1985control}
D.~J. Tannor and S.~A. Rice, J. Chem. Phys. \textbf{83}, 5013 (1985).

\bibitem{keay1995dynamic}
B.~J. Keay  \emph{et~al.}, Phys. Rev. Lett. \textbf{75}, 4102 (1995).


\bibitem{eckardt2005superfluid}
A.~Eckardt, C.~Weiss  and M.~Holthaus, Phys. Rev. Lett. \textbf{95}, 260404
  (2005).


\bibitem{lignier2007dynamical}
H.~Lignier  \emph{et~al.}, Phys. Rev. Lett. \textbf{99}, 220403 (2007).


\bibitem{kierig2008single}
E.~Kierig  \emph{et~al.}, Phys. Rev. Lett. \textbf{100}, 190405 (2008).


\bibitem{sias2008observation}
C.~Sias  \emph{et~al.}, Phys. Rev. Lett. \textbf{100}, 040404 (2008).


\bibitem{eckardt2010frustrated}
A.~Eckardt  \emph{et~al.}, Europhys. Lett. \textbf{89}, 10010 (2010).

\bibitem{Struck11}
J.~Struck  \emph{et~al.}, arXiv:1103.5944 .

\bibitem{schori2004excitations}
C.~Schori  \emph{et~al.}, Phys. Rev. Lett. \textbf{93}, 240402 (2004).


\bibitem{greif2010probing}
D.~Greif  \emph{et~al.}, Phys. Rev. Lett. \textbf{106}, 145302 (2011).

\bibitem{salger2009directed}
T.~Salger  \emph{et~al.}, Science \textbf{326}, 1241 (2009).

\bibitem{teichmann2009fractional}
N.~Teichmann, M.~Esmann  and C.~Weiss, Phys. Rev. A \textbf{79}, 063620 (2009).

\bibitem{esmann2011fractional}
M.~Esmann, N.~Teichmann  and C.~Weiss, Arxiv:1101.5272  (2011).

\bibitem{simon2011quantum}
J.~Simon  \emph{et~al.}, Arxiv:1103.1372  (2011).

\bibitem{fölling2007direct}
S.~F{\"o}lling  \emph{et~al.}, Nature \textbf{448}, 1029 (2007).

\bibitem{duan2003controlling}
L.~M. Duan, E.~Demler  and M.~D. Lukin, Phys. Rev. Lett. \textbf{91}, 090402
  (2003).


\bibitem{garcía2003spin}
J.~J. Garc{\'i}a-Ripoll and J.~I. Cirac, New J. Phys. \textbf{5}, 76 (2003).

\bibitem{kuklov2003counterflow}
A.~B. Kuklov and B.~V. Svistunov, Phys. Rev. Lett. \textbf{90}, 100401 (2003).


\bibitem{altman2003phase}
E.~Altman, W.~Hofstetter, E.~Demler  and M.~D. Lukin, New J. Phys. \textbf{5},
  113 (2003).

\bibitem{lewenstein2007ultracold}
M.~Lewenstein  \emph{et~al.}, Adv. Phys. \textbf{56}, 243 (2007).

\bibitem{fisher1989boson}
M.~P.~A. Fisher, P.~B. Weichman, G.~Grinstein  and D.~S. Fisher, Phys. Rev. B
  \textbf{40}, 546 (1989).

\bibitem{jaksch1998cold}
D.~Jaksch  \emph{et~al.}, Phys. Rev. Lett. \textbf{81}, 3108 (1998).


\bibitem{greiner2002quantum}
M.~Greiner  \emph{et~al.}, Nature \textbf{415}, 39 (2002).

\bibitem{widera2005coherent}
A.~Widera  \emph{et~al.}, Phys. Rev. Lett. \textbf{95}, 190405 (2005).


\bibitem{gerbier2006resonant}
F.~Gerbier  \emph{et~al.}, Phys. Rev. A \textbf{73}, 041602 (2006).

\bibitem{sebby2006lattice}
J.~Sebby-Strabley, M.~Anderlini, P.~S. Jessen  and J.~V. Porto, Phys. Rev. A
  \textbf{73}, 033605 (2006).

\bibitem{zhao1996rabi}
X.~G. Zhao, G.~A. Georgakis  and Q.~Niu, Phys. Rev. B \textbf{54}, 5235 (1996).

\bibitem{greiner2001exploring}
M.~Greiner  \emph{et~al.}, Phys. Rev. Lett. \textbf{87}, 160405 (2001).


\bibitem{hänggi1998driven}
M.~Grifoni and P.~H{\"a}nggi, Phys. Rep. \textbf{304}, 229 (1998).

\bibitem{winkler2006repulsively}
K.~Winkler  \emph{et~al.}, Nature \textbf{441}, 853 (2006).

\bibitem{auerbach1994interacting}
A.~Auerbach, \emph{{Interacting electrons and quantum magnetism}}.
\newblock Springer (1994).

\bibitem{trotzky2008time}
S.~Trotzky  \emph{et~al.}, Science \textbf{319}, 295 (2008).

\bibitem{Barmettler09}
P.~Barmettler  \emph{et~al.}, Phys. Rev. Lett. \textbf{102}, 130603 (2009).

\bibitem{Barthel09}
T.~Barthel, C.~Kasztelan, I.~P. McCulloch  and U.~Schollw\"ock, Phys. Rev. A
  \textbf{79}, 053627 (2009).

\bibitem{van2002interisotope}
E.~G. M.~V. Kempen, S.~Kokkelmans, D.~J. Heinzen  and B.~J. Verhaar, Phys. Rev.
  Lett. \textbf{88}, 093201 (2002).



\end{thebibliography}

\begin{thebibliography}{12}
\providecommand{\url}[1]{\texttt{#1}}
\providecommand{\urlprefix}{URL }
\providecommand{\eprint}[2][]{\url{#2}}

\bibitem[A1]{gerbier2006probing}
F.~Gerbier  \emph{et~al.}, Phys. Rev. Lett. \textbf{96}, 090401 (2006).

\end{thebibliography}
\end{document}